\documentstyle[aps,epsfig,prl,preprint]{revtex}

\newcommand{\csq}{\mbox{$\chi^{2}$}}

\newcommand{\ptpt}{\mbox{$p^{2}_{t}$}}
\newcommand{\pip}{\mbox{$\pi^{+}$}}
\newcommand{\pim}{\mbox{$\pi^{-}$}}
\newcommand{\pio}{\mbox{$\pi^{0}$}}

\newcommand{\ks}{\mbox{$K_{S}$}}
\newcommand{\kl}{\mbox{$K_{L}$}}

\newcommand{\kst}{\mbox{$K^{*}$}}
\newcommand{\kste}{\mbox{$K^{*}(892)$}}
\newcommand{\ksttw}{\mbox{$K_{1}(1270)$}}
\newcommand{\kstfh}{\mbox{$K_{1}(1400)$}}
\newcommand{\kstf}{\mbox{$K^{*}(1410)$}}
\newcommand{\kstft}{\mbox{$K^{*}_{2}(1430)$}}
\newcommand{\kstsngl}{\mbox{$\ks\pio$}}
\newcommand{\kstdbl}{\mbox{$\kste\pio$}}
\newcommand{\kstpair}{\mbox{(\ksttw -\kstfh )}}

\newcommand{\mev}{\mbox{MeV/c$^{2}$}}
\newcommand{\gev}{\mbox{GeV/c$^{2}$}}

\newcommand{\gamkste}{\mbox{$\Gamma_{r}(\kste )$}}
\newcommand{\gamf}{\mbox{$\Gamma_{r}(\kstf )$}}
\newcommand{\gamft}{\mbox{$\Gamma_{r}(\kstft )$}}
\newcommand{\gamtw}{\mbox{$\Gamma_{r}(\ksttw )$}}
\newcommand{\gamfh}{\mbox{$\Gamma_{r}(\kstfh )$}}

\begin{document}
\tightenlines

\begin{flushright}
RUTGERS-01-88\\
\end{flushright}

\begin{center}
\LARGE
{\bf Radiative Decay Width Measurements of Neutral Kaon 
Excitations Using the Primakoff Effect}\\
\normalsize
\vspace*{.2 in}

\parindent=0.in
\parskip 0 in
A.~Alavi-Harati$^{12}$,
T.~Alexopoulos$^{12}$,
M.~Arenton$^{11}$,
K.~Arisaka$^2$,
S.~Averitte$^{10}$,
R.F.~Barbosa$^{7,**}$,
A.R.~Barker$^5$,
M.~Barrio$^4$,
L.~Bellantoni$^7$,
A.~Bellavance$^9$,
J.~Belz$^{10}$,
D.R.~Bergman$^{10}$,
E.~Blucher$^4$, 
G.J.~Bock$^7$,
C.~Bown$^4$, 
S.~Bright$^4$,
E.~Cheu$^1$,
S.~Childress$^7$,
R.~Coleman$^7$,
M.D.~Corcoran$^9$,
G.~Corti$^{11}$, 
B.~Cox$^{11}$,
A.~Cunha$^{10}$,
A.R.~Erwin$^{12}$,
R.~Ford$^7$,
A.~Glazov$^4$,
A.~Golossanov$^{11}$,
G.~Graham$^{4}$, 
J.~Graham$^4$,
E.~Halkiadakis$^{10}$,
J.~Hamm$^1$,
K.~Hanagaki$^{8}$,  
S.~Hidaka$^8$,
Y.B.~Hsiung$^7$,
V.~Jejer$^{11}$,
D.A.~Jensen$^7$,
R.~Kessler$^4$,
H.G.E.~Kobrak$^{3}$,
J.~LaDue$^5$,
A.~Lath$^{10}$,
A.~Ledovskoy$^{11}$,
P.L.~McBride$^7$,
D.~Medvigy$^{10}$,
P.~Mikelsons$^5$,
E.~Monnier$^{4,*}$,
T.~Nakaya$^{7}$,
K.S.~Nelson$^{11}$,
H.~Nguyen$^7$,
V.~O'Dell$^7$, 
R.~Pordes$^7$,
V.~Prasad$^4$,
X.R.~Qi$^7$, 
B.~Quinn$^4$,
E.J.~Ramberg$^7$, 
R.E.~Ray$^7$,
A.~Roodman$^{4}$,  
S.~Schnetzer$^{10}$,
K.~Senyo$^{8}$, 
P.~Shanahan$^7$,
P.S.~Shawhan$^{4}$,
J.~Shields$^{11}$,
W.~Slater$^2$,
N.~Solomey$^4$,
S.V.~Somalwar$^{10,\dagger}$, 
R.L.~Stone$^{10}$, 
E.C.~Swallow$^{4,6}$,
S.A.~Taegar$^1$,
R.J.~Tesarek$^{10}$, 
G.B.~Thomson$^{10}$,
P.A.~Toale$^5$,
A.~Tripathi$^2$,
R.~Tschirhart$^7$,
S.E.~Turner$^2$, 
Y.W.~Wah$^4$,
J.~Wang$^1$,
H.B.~White$^7$, 
J.~Whitmore$^7$,
B.~Winstein$^4$, 
R.~Winston$^4$, 
T.~Yamanaka$^8$,
E.D.~Zimmerman$^{4}$
\vspace*{.1 in} 
\footnotesize

$^1$ University of Arizona, Tucson, Arizona 85721 \\
$^2$ University of California at Los Angeles, Los Angeles, California 90095 \\
$^{3}$ University of California at San Diego, La Jolla, California 92093 \\
$^4$ The Enrico Fermi Institute, The University of Chicago, 
Chicago, Illinois 60637 \\
$^5$ University of Colorado, Boulder, Colorado 80309 \\
$^6$ Elmhurst College, Elmhurst, Illinois 60126 \\
$^7$ Fermi National Accelerator Laboratory, Batavia, Illinois 60510 \\
$^8$ Osaka University, Toyonaka, Osaka 560-0043 Japan \\
$^9$ Rice University, Houston, Texas 77005 \\
$^{10}$ Rutgers University, Piscataway, New Jersey 08854 \\
$^{11}$ The Department of Physics and Institute of Nuclear and 
Particle Physics, University of Virginia, 
Charlottesville, Virginia 22901 \\
$^{12}$ University of Wisconsin, Madison, Wisconsin 53706 \\
$^{*}$ Permanent address C.P.P. Marseille/C.N.R.S., France \\
$^{**}$Permanent address University of S\~{a}o Paulo, S\~{a}o Paulo, Brazil \\
\vspace*{.1 in}
$^{\dagger}$ To whom correspondence should be addressed. Electronic
address: somalwar@physics.rutgers.edu\\
\vspace*{.1 in}

\centerline{ \bf The KTeV Collaboration}

\normalsize


\end{center}
\normalsize

\newpage
\begin{abstract}

We produce 
a sample consisting of 147 candidate events, with minimal backgrounds, of
the mixed axial vector pair \kstpair\ by exciting
\kl 's in the Coulomb field of lead and report the first measurements
of the radiative widths 
$\Gamma_{r}(\kstfh )=280.8\pm 23.2(stat)\pm 40.4(syst)$~keV
and $\Gamma_{r}(\ksttw) = 73.2\pm
6.1(stat)\pm 28.3(syst)$ keV.  
We also place 90\% CL upper limits 
$\Gamma_{r}(\kstf ) \leq 52.9$ keV for the vector state
and $\Gamma_{r}(\kstft ) \leq 5.4$ keV for the tensor state.  
These measurements allow for significant tests of quark-model
predictions of radiative widths for the low-lying vector mesons.

\end{abstract}

\vspace*{.2 in}

\begin{figure}
\begin{center}
   \parbox{3.0in}{\epsfxsize=3.0in\epsffile
{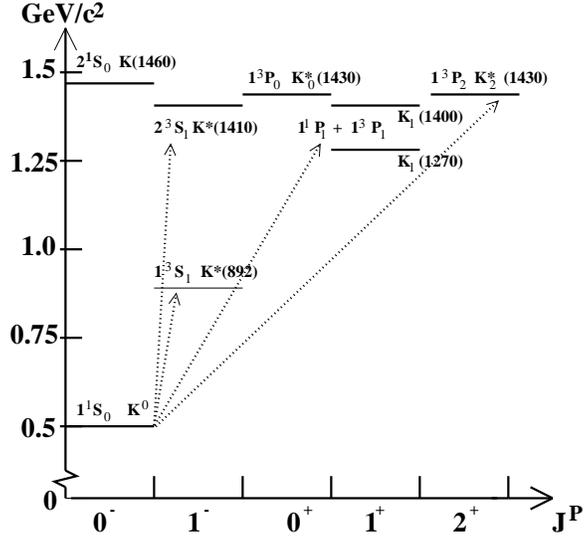}}
\end{center}
\caption{Mass vs. angular momentum and parity ($J^{P}$) 
for neutral kaon resonances. 
Arrows indicate resonances accessible by Primakoff excitation. }
\label{fig:level}
\end{figure}

Several resonant excitations of the neutral kaon are known to
exist~\cite{pdg}, most having been observed {\em indirectly}
using partial wave analysis~\cite{daum}.  Figure~\ref{fig:level} is a
schematic representation of the neutral kaon excitations with central masses
less than 1.5~\gev. The axial vector pair \kstpair\ is interesting because
it is a (coherent) mixture of the singlet~$^1P_1$ and the triplet~$^3P_1$ 
states~\cite{SLAC}, parameterized by the mixing angle $\Theta$:
\mbox{\(  \ksttw = -~^3P_1\cdot\sin\Theta   +~^1P_1\cdot\cos\Theta \)}
and 
\mbox{\(  \kstfh = ~^3P_1 \cdot\cos\Theta   +~^1P_1\cdot\sin\Theta \).}
The radiative decay widths of the kaon excitations,
$\Gamma_{r}(\kst ) =\Gamma (\kst\rightarrow K+\gamma)$, are
sensitive to the magnetic moments of the constituent
quarks~\cite{becchi}. Radiative widths have
been calculated for low-lying mesons using both a dynamic quark
model~\cite{barikdash} and a relativistic quark model~\cite{daspanda,GodIs}.
Experimentally, only \gamkste\ has been measured~\cite{Carlsmith} so far.

The Primakoff effect~\cite{primakoff}, i.e. excitation by the Coulomb
field, can be used to measure radiative widths since it is the inverse
of radiative decay.  In this Letter, we use the full dataset collected
during the 1996-97 run of the KTeV experiment at Fermilab to study
Primakoff production in two channels: the six-body \kstdbl\ channel,
exemplified by \kstf\ or \kstfh\ $\rightarrow\kste\pio\rightarrow
[\ks\pio ]\pio \rightarrow [(\pip\pim )(\gamma\gamma )](\gamma\gamma
)$, which has two \pio's, and the four-body \kstsngl\ channel, exemplified by
\kste\ or \kstf\ $\rightarrow\ks\pio \rightarrow (\pip\pim
)(\gamma\gamma )$, which has a single \pio . In the \kstdbl\
channel, we observe 147 candidate events which are predominantly the
axial vector \kstfh\ with a small admixture of \ksttw . Using a large
sample of \kste 's from the \kstsngl\ channel for normalization, we
report the first measurements of the radiative widths for the axial
vector pair.  We also use the \kstsngl\ channel to place the first
upper limit on \gamf\ and a stringent upper limit on \gamft .

For high particle energies and small production angles,
the rate of exciting a \kl\ to a \kst\ in the Coulomb field of a nucleus
A is given by~\cite{berlad}
\begin{equation}
\frac{d\sigma}{dt}(\kl +A\rightarrow\kst +A) 
= \pi \alpha Z^{2} \left(\frac{2S_{K^{*}}+1}
{2S_{K}+1} \right) \frac{\Gamma (K^{*} \rightarrow K+\gamma)}{k^{3}}
\frac{t'}{t^{2}} |f_{EM}|^{2} , \label{eq:thxsect}
\end{equation}
where $\alpha$ is the fine structure constant, 
$Z$ is the atomic number of the nucleus, 
$S_{K}$ and $S_{K^{*}}$ are the spins of \kl\ and the resonance,
respectively, 
$k=(m_{K^{*}}^{2}-m_{K}^{2})/2m_{K^{*}}$, $t$ 
is the magnitude of the square of the momentum transfer
and $t'=t-t_{min}$, 
$\sqrt{t_{min}}=(m_{K^{*}}^{2}-m_{K}^{2})/2P_{K}$, where 
$P_{K}$ is the laboratory momentum of the \kl . 
Finally, $f_{EM}$ is the nuclear electric form factor.
Thus, the rate of Primakoff production is directly proportional to the 
radiative width.

KTeV utilized an $800$ GeV/c proton beam to generate two neutral beams
consisting kaons, neutrons and some hyperons.  
In the E832 configuration~\cite{eprime}, one of the beams passed through a
regenerator which was located \mbox{$\sim$124~m} from the target. The
regenerator consisted of 84 modules of 2cm-thick plastic scintillator
followed by a module composed of a lead-scintillator sandwich.  Since the
Primakoff effect is proportional to $Z^2$ of the target material, more
than \mbox{98\% } of the observed Primakoff excitations 
(equation~\ref{eq:thxsect}) were produced in the 
final lead pieces.
The regenerator was instrumented with photomultiplier tubes which
enabled us to tag and reject backgrounds from inelastic interactions.
We detect \pip\pim\ tracks from \ks\ decays using a drift chamber 
spectrometer system and photons from \pio\ decays
using a pure CsI electromagnetic calorimeter.  The event trigger was
initiated by signals from two scintillator hodoscopes located
downstream of the spectrometer and required hits in the drift chambers
consistent with two oppositely charged tracks.  The decay volume was 
surrounded by a near-hermetic set of devices to veto photons. 

In the offline analysis, the fiducial region for the decay vertex of
$\ks\rightarrow\pip\pim$ is restricted to 15~m downstream of the
regenerator.  We reconstruct \pio 's using pairs of energy clusters in
the calorimeter.  The clusters are required to have energies greater
than 1 GeV and photon-like spatial distributions. To reject electrons,
we require that the ratio of energy deposited in the calorimeter to
the particle momentum as measured by the spectrometer be $<0.8$.

\begin{figure}
\begin{center}
   \parbox{3.0in}{\epsfxsize=3.0in\epsffile
{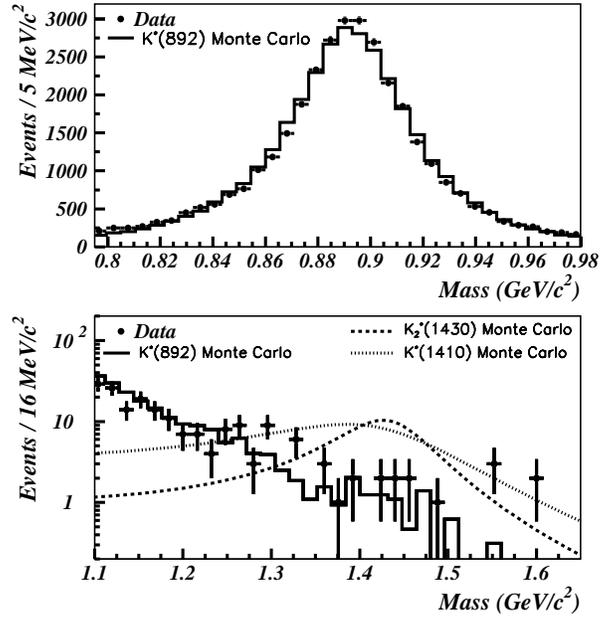}}
\end{center}
\caption{
Top: \ks\pio\ invariant mass in the four-body channel showing 
$\kste\rightarrow\kstsngl$ decays.
Bottom: The same \ks\pio\ invariant mass in the 1.4 \gev\ region. 
\kstft\ and \kstf\ simulations are also shown to arbitrary scale.
No \kstft\ or \kstf\ resonance is apparent. 
} 
\label{fig:892_himass}
\end{figure}

To reconstruct the $\kste\rightarrow\kstsngl$ decays used
for normalization, in the four-body channel we require the invariant 
masses of the
$\gamma\gamma$ and the $\pip\pim$ to be within 10 \mev\ of the \pio\
and \ks\ invariant masses, respectively.  We isolate Primakoff
(forward) production by demanding that the square of the
transverse momentum (\ptpt ) of the $\pip\pim\gamma\gamma$ with respect
to a line connecting the target and the decay vertex of \kste\ be 
less than $0.001~(GeV/c)^2$.  We further require \pip\pim\ 
\ptpt\ $>0.01~(GeV/c)^2$
because the {\em daughter} \ks\ recoils against the \pio . The resulting
sample of 29,399
$\kste\rightarrow\ks\pio$ decays with \ks\ energy between 30 and
210~GeV, and the \kste\ energy between 55 and 225~GeV is shown in
figure~\ref{fig:892_himass}(top).

\begin{figure}
\begin{center}
   \parbox{3.0in}{\epsfxsize=3.0in\epsffile
{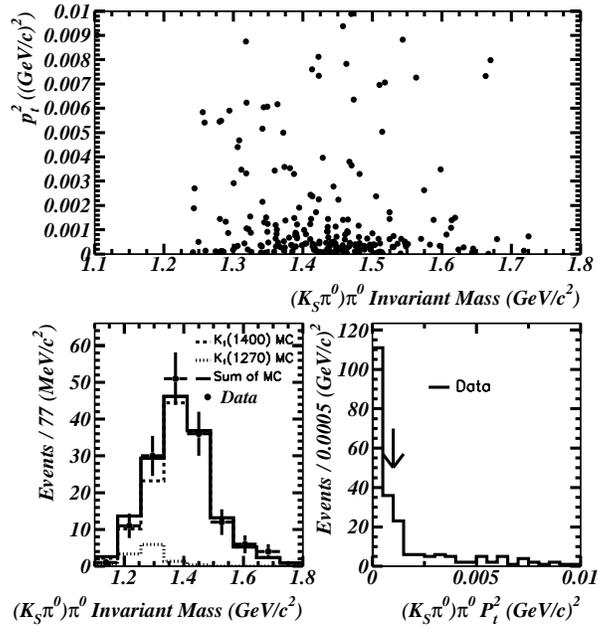}}
\end{center}
\caption{ 
Top: (\kste\pio ) \ptpt\ vs. invariant mass after all other cuts.
Bottom Left: Projection onto the abscissa after 
the \ptpt\ cut.  Decomposition of the observed signal into 
\ksttw\ and \kstfh\ is also shown. Bottom Right: Projection onto the ordinate
after the mass cut. Note the sharply forward nature of Primakoff production.}
\label{fig:1400_mpt_ovlay}
\end{figure}

\begin{figure}
\begin{center}
   \parbox{3.0in}{\epsfxsize=3.0in\epsffile
{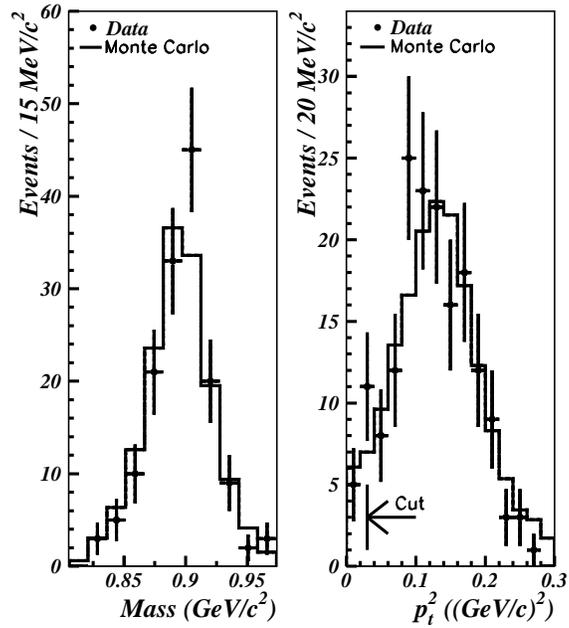}}
\end{center}
\caption{Data/MC comparisons for 
the \ks\pio\ invariant mass (left) and
the \ptpt\ for the observed $\kstpair\rightarrow\kste\pio$ signal (right).
A Jacobian distribution in \ptpt\ indicates the recoil of the daughter \kste\ 
against the $\pi^0$.
We discard events to the left of the arrow.
}
\label{fig:1400_m_p892datMC}
\end{figure}

The requirements for the \kstdbl\ six-body channel are similar, 
except for changes to account for the extra \pio\ and differences in 
kinematics. The photon pairings for the two \pio 's
are determined using a \csq\ formed by comparing 
$M_{\gamma\gamma }$ and $M_{\pip\pim\gamma\gamma }$ to
the known masses of \pio\ and \kste , respectively.
The \ks\pio\ mass for the {\em daughter} \kste\ is required to be within 101 \mev\ 
(two mass-widths) of the \kste\ mass and its \ptpt\ to be $>0.03~(GeV/c)^2$.
The \ptpt\ cut also serves to eliminate background from Primakoff-produced \kste 's
when accompanied by an accidental  \pio .  
To eliminate events in which two kaons decay to a charged and a 
neutral pion pair,
we remove events for which the four-photon 
invariant mass is within 20 MeV of the \kl\ mass.
The resulting sample of (\ks\pio )\pio\ events with total energy greater than 90~GeV
is depicted in figure~\ref{fig:1400_mpt_ovlay} and shows a clustering near 
1.4 \gev . The mass projection shows the resonant signature exhibited by
events with $\ptpt <0.001~(GeV/c)^{2}$ and the \ptpt\ projection shows 
the sharp fall-off confirming Primakoff production.
There are 147 events within the 
mass fiducial region (1.1-1.64 \gev ).
Figure~\ref{fig:1400_m_p892datMC} shows
the invariant mass and \ptpt\ of the daughter \kste\ where
the \ptpt\ displays a Jacobian distribution expected 
of a daughter particle in a two-body decay.

The possible candidates for the observed \kstdbl\ resonance are
$K^{*}_{0}(1430)$, K(1460), \kstf, \kstft , \ksttw , and \kstfh\
(figure~\ref{fig:level}).  The selectivity of the Primakoff effect
rules out $K^{*}_{0}(1430)$ and K(1460) because of
spin-parity conservation and the J=0$\not\rightarrow$ J=0 selection
rule, respectively.
Contributions from the vector \kstf\ and tensor \kstft\ can be
eliminated because both have significant branching fractions to
\mbox{\kstsngl ~\cite{pdg}}, yet we see no evidence for their presence in
this (\kstsngl ) channel; note the lack of resonance near 1.4 \gev\ in
figure~\ref{fig:892_himass} (bottom).  We fit a combination of \kste\ and
\kstf (\kstft ) simulations to the data and
confirm that the signal from \kstf (\kstft ) in the \kstsngl\
channel is consistent with zero: $4.0\pm 6.0 (0.1\pm 3.8)$
\kstf (\kstft ) events. Using the known branching
fractions~\cite{pdg} of \kstf\ and \kstft\  to \kstsngl\ and \kstdbl ,
we translate these results into 
a negligible $2.4\pm 3.6 (0.0\pm 0.7)$ event contribution 
of \kstf (\kstft ) in the \kstdbl\ channel. Thus we are left
with only the axial vector pair \kstpair\ as a possible
candidate for the observed resonance in the \kstdbl\ channel.

We cross-checked the axial vector nature of the observerd signal using
the distributions of Gottfried-Jackson (GJ) angles\footnote[2]{In the
GJ frame the excited kaon is at rest at the origin, the z-axis is
along the momentum of the incoming \kl\ and the y-axis is
perpendicular to the production plane~\cite{GJ}.}  $\theta$ and
$\phi$.  These distributions generally confirm our axial vector
assignment. However, due to the relatively strong angular dependence of the
detector acceptance, they do not have strong discrimination power
between the axial vector pair and \kstf\ and \kstft .
%

We now compute the radiative widths for the axial vector pair.  It is
difficult to decompose the mass spectrum of the observed
signal into \ksttw\ and \kstfh\ because their mass separation is
comparable to their widths.  Nonetheless, mass information alone
tells us that the contribution from \ksttw\ is slight: only $8.8\pm 8.6$
events are due to \ksttw .  However, 
a significantly better resolution is possible because the Primakoff effect 
can produce only the singlet ($^1P_1$) component of the axial vector pair~\cite{br}
and the singlet-triplet mixing angle $\Theta$
has been measured. Using $\Theta = 56\pm 3^{\circ}$~\cite{daum}
together with the known branching ratios of \ksttw\ and \kstfh , 
we resolve the observed signal into $11.4\pm 1.0(stat)
\pm 4.1(ext\ syst)$ \ksttw\ events and $134.4\pm 11.1(stat) \mp
4.1(ext\ syst)$ \kstfh\ events, where the (external) systematic error
is due to the measurement uncertainties in the mixing parameter
$\Theta$ and in the \ksttw\ and \kstfh\ branching fractions to the
\kstdbl\ channel. This decomposition, depicted in
figure~\ref{fig:1400_mpt_ovlay}, leads to 
$\gamtw = 73.2\pm 6.1
(stat)\pm 8.2(int\ syst) \pm 27.0 (ext\ syst)$ keV and $\gamfh =
280.8\pm 23.2(stat)\pm 31.4(int\ syst)\pm 25.4(ext\ syst)$ keV,
where we have used our \kste\ sample (figure~\ref{fig:892_himass}) 
for normalization purposes since \gamkste\ is known 
experimentally~\cite{Carlsmith}. Our measurements share internal
systematic errors of 8.7\% due to uncertainties in the strong
production (discussed below), 6.6\% due to detector acceptance effects, 
and 2.4\% due to the 3.6 event uncertainty in
the possible contributions from \kstf\ and \kstft , as discussed earlier.  
The uncertainty in the \kste\ radiative width 
measurement~\cite{Carlsmith} causes 
an additional 8.5\% (external) systematic error. 

Primakoff production is 
characterized by a sharp ($\sim t^{-1}$) forward 
production \mbox{(equation~\ref{eq:thxsect})} allowing a strict
$\ptpt <0.001 (GeV/c)^{2}$ cut which virtually eliminates all potential backgrounds;
see figure~\ref{fig:1400_mpt_ovlay}.
Based on an extrapolation from the large \ptpt\ ($>0.1~(GeV/c)^{2}$) region, 
we estimate 1.2 events out of 147 signal candidate events to be due to 
incoherent production and other possible backgrounds such as those from 
the decay products of the $\Lambda$'s and $\phi$'s produced when neutrons in the 
beam interact with the regenerator.

Coherent strong production and its interference (with unknown
strength) with Primakoff production are expected to be small at our
energies~\cite{Carlsmith}.  Indeed,
a maximum likelihood fit in the \ptpt\ variable for
the strong production and the strength of the strong-Coulomb
interference using the prescription given in~\cite{Carlsmith,strong}
indicates that the strength of interference preferred by our data
is consistent with zero.
A constructive (destructive) interference 
would mean that the actual number of Primakoff events
is less (more) than what we observe.
The mean change in our estimate of 
Primakoff production corresponding to 
one standard deviation variation in the interference strength
is 8.7\%, which we have taken to be the systematic 
error due to the uncertainties in strong production.

Earlier, we used the absence of a resonance in the \kstsngl\ 
channel at $\sim$1.4
\gev (figure~\ref{fig:892_himass}) to limit the \kstf\ and \kstft\ 
contributions to the observed \kstpair\ axial vector pair signal.  A further
benefit of this finding is that we are able to limit
the radiative widths \gamf\ and \gamft\ to 
52.9 and 5.4 keV,
respectively, at 90\% CL.  \gamf\ has not been examined experimentally before, 
whereas \gamft\ was previously limited to 84~keV at 90\% CL~\cite{Carlsmith}.
The  \gamft\ limit is far stricter than the \gamf\ limit
principally because the branching fraction for
$K^{*}_{2}(1430) \rightarrow K_{S}\pi^{0}$ is substantially larger~\cite{pdg} 
than the same for $K^{*}(1410) \rightarrow K_{S}\pi^{0}$. 

The predicted radiative widths for the axial vector mesons~\cite{russians}, 
are  
538 keV for \kstfh\ and 175 keV for \ksttw ; compare to our results,
$280.8\pm 46.6$~keV and $73.2\pm 28.9$~keV, respectively.
We note that the theoretical model in~\cite{russians}
is very sensitive to the quark masses ($m_{u,s}$) and rms
momenta ($\beta_{uu,us,ss}$) of the quarks within the mesons. The 
predictions are based on certain choices for $m$ and $\beta$, but
other choices with up to 30\% 
variation in $m$ are possible.

Our 90\% CL upper limit on the vector \kstf\ radiative width is 
52.9
keV.  In the naive quark model, this state is the first radial
excitation of \kste\ and its radiative width calculation should be
similar to that for \kste , for which 
$\gamkste = 116.5\pm 9.9$ keV~\cite{Carlsmith}.  
The smaller value for \kstf\ may be due to a
reduced overlap of the quark wavefunction for this higher radial
excitation, but further guidance from theory is needed.

Finally, we have substantially improved the upper limit on the radiative 
width of the tensor \kstft\ from 84~keV~\cite{Carlsmith} to 
5.4~keV (at 90\% CL). 
Babcock and Rosner~\cite{br} used $SU(3)$ invariance to predict
that excitations with $J^{PC}=1^{++}$
or $2^{++}$ would have vanishing radiative widths. In the limit of
$SU(3)$, \kstft\ has $C=+1$; thus, our 
limit lends support to Babcock and Rosner's prediction~\cite{br} and
serves as a direct test of the naive quark
model and $SU(3)$-breaking.

We thank J. Bronzan 
for theoretical guidance and
gratefully acknowledge the support and effort of the Fermilab
staff and the technical staffs of the participating institutions for
their vital contributions.  This work was supported in part by the U.S. 
Department of Energy, The National Science Foundation and The Ministry of
Education and Science of Japan. 
In addition, A.R.B., E.B. and S.V.S. 
acknowledge support from the NYI program of the NSF; A.R.B. and E.B. from 
the Alfred P. Sloan Foundation; E.B. from the OJI program of the DOE; 
K.H., T.N. and M.S. from the Japan Society for the Promotion of
Science; and R.F.B. from the Funda\c{c}\~{a}o de Amparo \`{a} 
Pesquisa do Estado de S\~{a}o Paulo.  P.S.S. 
acknowledges receipt of a Grainger Fellowship.

\end{document}